\documentstyle[preprint,aps]{revtex}

\begin{document}

\draft

\title{Mixed-state penetration depths in s-wave and d-wave superconductors} 

\author{Yong Wang and A. H. MacDonald} 

\address{Department of Physics, Indiana University, 
Bloomington, IN 47405} 

\date{\today} 

\preprint{IUCM96-020} 

\maketitle 

\begin{abstract} 

We report on a microscopic calculation of the current and magnetic field
distribution in the mixed state of model s-wave and d-wave
superconductors.  For the d-wave case, we find that 
corrections to London theory are important at fields 
small compared to $H_{c2}$.   The field distribution 
is influenced by both superfluid-velocity dependence and 
non-locality in the current response function of the mixed state.  
We compare our calculations with recent muon spin rotation
measurements in high temperature superconductors. 

\end{abstract} 


The Meissner effect, in which weak external magnetic 
fields are exponentially screened from the bulk, 
is a basic property of superconducting metals. 
It is a consequence of the linear relationship between 
circulating currents (${\bf J}$) and the magnetic flux density (${\bf h}$)
in a superconductor expressed by the London equation: 
\begin{equation}
{\bf h} + \frac{4 \pi \lambda^2 }{c} \nabla \times {\bf J} = 0.
\label{eq:london}
\end{equation}
In Eq.(\ref{eq:london}) $\lambda$, the penetration
depth, determines the spatial extent of magnetic fields
near the surface of a superconductor in the Meissner state.
Measurements of this parameter 
in a particular material provide important information on the 
microscopic nature of its superconducting state.  
For example, recent measurements\cite{hardy,kiefl} of the temperature
dependence of $\lambda$, which microscopic theory relates to the 
quasiparticle energy spectrum\cite{tinkham},
have provided important evidence for d-wave pairing\cite{dwave}
in high-temperature superconductors.  

Both nuclear magnetic resonance (NMR) and muon spin rotation ($\mu$SR) 
measurements
of $\lambda$ are based on the relationship between 
magnetic field inhomogeneity in the mixed state of
a type-II superconductor and the penetration depth.
In the mixed state, Eq.(\ref{eq:london})
applies only outside the vortex cores and the London equation 
becomes,
\begin{equation}
{\bf h} + \frac{4 \pi \lambda^2}{c} \nabla \times {\bf J}  
= \hat {\bf z} \Phi_0 \sum_i \delta^{(2)}({\bf r} - {\bf R}_i), 
\label{eq:londonmixed}
\end{equation} 
where ${\bf R}_i$ is a vortex position and $\Phi_0$ is a 
superconducting flux quantum.   
The right hand side of Eq.(\ref{eq:londonmixed}) accounts for
the phase winding of the superconducting order 
parameter around each vortex.  The magnetic field distribution
inside the superconductor can be evaluated\cite{kossler}
from Eq.(\ref{eq:londonmixed})
and the Maxwell equation, $ \nabla \times {\bf h} = (4 \pi /c) {\bf J}$,
once the vortex positions are specified.  
Eq.(\ref{eq:londonmixed}) should be valid  
provided that: i) the magnetic field is sufficiently weak compared to the 
upper critical field so that vortex cores occupy a small fraction 
of the volume; ii) the distance scales on which 
currents and fields vary inside the superconductor are larger
than the non-locality length of the current response kernel; iii) and 
the fields and currents are small enough to be in the linear
response regime.  This last condition is especially 
pertinent in the case of d-wave superconductors since 
nodes in the energy gap lead to large non-linearities\cite{sauls,stojkovic} 
in the current response, even at low temperatures.  At present 
there is no detailed theory which accounts for non-linear
supercurrent response in the inhomgeneous mixed state.  In this 
article we report on exact microscopic mean-field-theory calculations
of the current and magnetic field distributions in the mixed 
state of model s-wave and d-wave superconductors.
We summarize our results in terms of an
effective penetration depth which would be inferred from the 
calculated field distribution under the assumption that
Eq.(\ref{eq:londonmixed}) applies locally.  We find that 
this effective penetration depth increases with increasing 
field, and hence increasing supercurrent densities, much
more strongly for d-wave superconductors than for s-wave 
superconductors and that the linear temperature dependence 
of $\lambda$ which signals $d$-wave superconductivity 
in the absence of a magnetic field saturates
at a temperature $k_B T^{*} \sim (H/H_{c2})^{1/2} \Delta_0$. 

Our numerical calculations start from self-consistent 
solutions\cite{gygi,dorsey,mcmaster,cincinatti}
of the mean-field Bogoliubov-de Gennes (BdG) equations\cite{deGennes,caroli}  
for generalized Hubbard models on stacked magnetically coupled square lattices.
We regard the Hubbard models
as plausible low-energy effective models for high-temperature
superconductors.  The class of Hubbard models we consider is paramaterized 
by the nearest neighbor hopping energy ($t$), the 
on site interaction energy ($U$) and the nearest neighbor
interaction energy ($V$) and the parameters can be chosen  
so as to obtain s-wave or d-wave superconductors at 
zero magnetic field. The model parameters for the calculations 
discussed below were chosen to give low temperature coherence lengths
comparable to estimates made for typical high temperature superconductors.
We have solved the BdG equations\cite{latticestructure} self-consistently
for s-wave and d-wave mixed states, 
assumming\cite{remark1} a constant magnetic flux density $H \hat z$,
as detailed in an earlier publication.\cite{wang1}
The current flowing along the bond of the lattice linking
site $i$ and site $j$ is related to the self-consistent BdG quasiparticle
amplitudes by
\begin{equation}
J_{ij}= \frac {4e}{\hbar ad} Im \Bigl \lbrace t_{ij} \sum_{\alpha}\bigl[ 
u_i^{\alpha \ast}u_j^{\alpha}f_{\alpha} + v_i^{\alpha}v_j^{\alpha \ast} 
(1-f_{\alpha})\bigr]\Bigr \rbrace ,        
\label{eq:linkcurrent}
\end{equation}
where $a$ is in-plane lattice constant,  
$d$ is the separation between square lattice layers, 
and $f_{\alpha}=[ \exp (E_{\alpha}/k_BT)+1]^{-1}$ is 
the Fermi function for quasiparticle excitations of 
energy $E_{\alpha}$. 
The spatial variation in the 
flux density $h$ in each plaquette of the square lattice 
is then evaluated from a discrete version of the magnetostatic  
Maxwell equation:
\begin{equation}
{\bf h}({\bf r})= \frac{4 \pi i}{c}
\sum_{\bf G \ne 0} \frac {{\bf G} \times {\bf J}({\bf G})} 
{G^2} \exp \bigl(i{\bf G} \cdot {\bf r} \bigr) + H .     
\label{eq:field}
\end{equation}

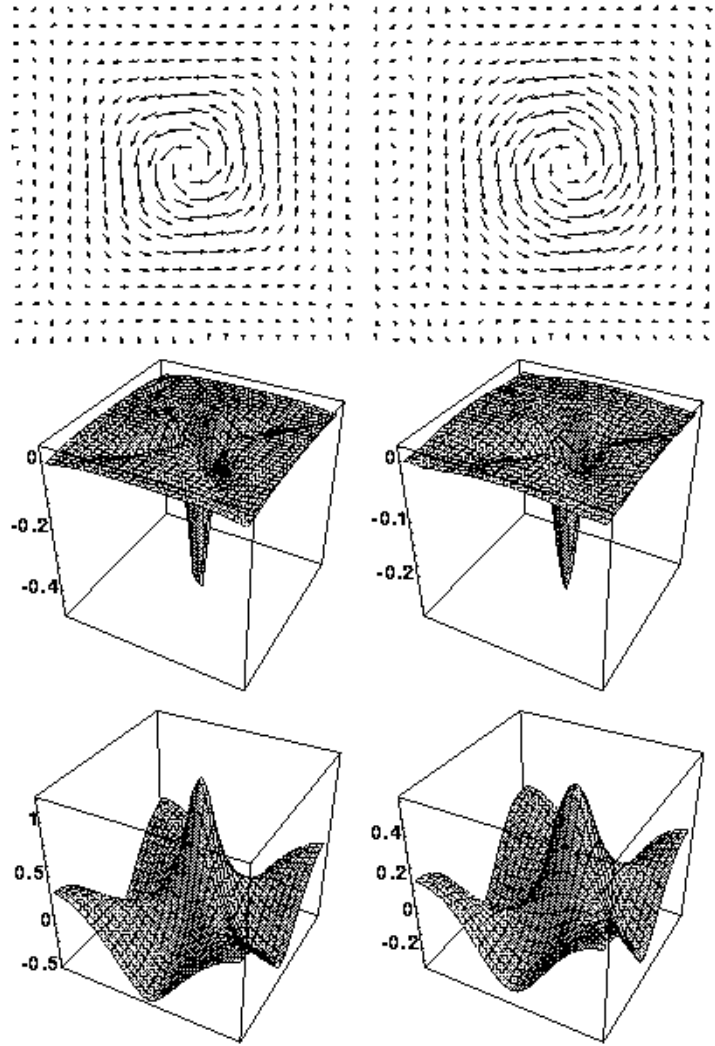
\begin{figure}
\caption{Top row: Current circulating around a vortex. Arrows represent
the directions of flow. The length of an arrow represents the magnitude of
the current at the site. Middle row: Local penetration depth as defined
in the text.  The minimum
in this quantity occurs at the center of the vortex core.
Bottom row: Magnetic field distribution.  The peak occurs at the
center of the vortex core.
The plotted quantity is the difference between total field and the spatial
average field
in arbitrary units. The applied field is $H=0.04H_{c2}$.
The figures in left and right columns are for
$s$-wave case and $d$-wave case respectively.}
\label{fig:one}
\end{figure}

Typical results are shown in Fig.~\ref{fig:one} for both s-wave
(left column) and d-wave (right column) superconductors.
In these figures we show the spatial dependence of the 
current density, the local magnetic flux density, 
and a local penetration depth which we define by 
\begin{equation}
\lambda^{-2}_{loc} \equiv -\frac {4 \pi}{c} 
\frac{\nabla \times {\bf J}}{h}. 
\label{eq:locallambda}
\end{equation} 
These results were obtained for $T=0$ 
at fields with approximately $900$ plaquettes 
per superconducting flux quantum which corresponds, for the model studied,
to $H\simeq 0.04 H_{c2}$ in both $s$-wave and $d$-wave cases.

The screening current circulating around the vortex core 
is largest in magnitude at the core edge. 
Qualitatively, this property can be easily understood in 
a Ginzburg-Landau type of analysis,  
where the current density is related to the order parameter 
$\psi=|\psi|\exp(i\phi)$ 
and the superfluid velocity ${\bf v}_s$ by  
\begin{equation}
{\bf J}=2e|\psi|^2{\bf v}_s
=\frac{2e}{m^{\ast}}|\psi|^2\Bigl(\hbar \nabla \phi 
-\frac{2e}{c}{\bf A}\Bigr).   
\label{current}
\end{equation}
Since $\phi$ winds by $2\pi$ around a vortex and $|\psi|$ vanishes linearly
at vortex center, $J \sim r$ near
the core center and $J \sim r^{-1}$ (for $r\ll \lambda$) outside the core. 
Our results show a distortion in the circular flow pattern generally
expected around the vortex core which we attribute\cite{remark2}
to a tendency for current to flow along the principle symmetry 
axes of the square lattice.  We find that the local penetration depth is
tolerably constant outside the vortex core\cite{remark3}, as assummed in the 
simple London theory, except 
where the current flow bends relative abruptly. 
Similar band structure effect has been seen in scanning-tunneling-microscopy 
experiments for conventional superconductors by Hess {\it et al.}\cite{hess}. 
However, near and inside the vortex core
$\lambda^{-2}_{loc}$ changes rapidly,  
presumably principally because the magnitude of the order parameter 
is changing rapidly. 
There is no qualitatively difference between $s$-wave and $d$-wave cases
apparent in the results at a particular temperature and field shown in 
Fig.~\ref{fig:one}. The deep minimum in the local penetration depth
in the vortex core is effectively replaced by a $\delta$-function in the 
London approximation.

In the Meissner state\cite{sauls,stojkovic} of a $d$-wave superconductor, 
the dependence of the penetration depth on superfluid velocity
leads to important non-linear terms in the current response and to
a field dependent effective penetration depth.  
In the mixed state, the dependence of the magnetic field 
on {\it two} spatial coordinates and the presence of vortex cores where
the order parameter magnitude is not constant  
complicate the analysis of non-linear effects.  In fact, the current
density in the mixed state is largest near the vortex cores and 
this complicated region might therefore be expected to
contribute importantly to the field dependence of any  
effective penetration depth, making the development of an 
analytic theory truly difficult.  The magnetic  
field at any point in the vortex lattice depends on the 
current distribution everywhere and hence its distribution
function depends on some complicated average of the non-local,
non-linear, and inhomogenity effects 
which appear in the local penetration depth.  
It is useful to define a magnetic-field-dependent effective 
penetration depth $\lambda$ in the mixed state by mimicing the 
analyses of $\mu$SR or NMR experiments which are based on the 
London equation.  This $\lambda$ may be thought of as 
an appropriate average of the local penetration depth discussed 
above. 

In a $\mu$SR experiment one measures the time evolution of the muon 
polarization vector precessing about the internal magnetic field. The 
frequency spectrum is proportional to the magnetic field probability 
distribution in the sample. 
When the London equation
is valid nearly everywhere it can be shown\cite{brandt} that 
$\lambda$ is related to the width of the magnetic field probability 
distribution 
\begin{equation} 
P(h)=\frac {1}{A}\int d\, {\bf r}\, \delta (h-h({\bf r}))  
\label{eq:prob}
\end{equation} 
by the following expression, 
\begin{equation}
\lambda^{-4}=C\, \langle \Delta h^2 \rangle 
\equiv C \int d\,h\,P(h)\,(h-H)^2,
\label{eq:width}
\end{equation} 
where the constant $C$ is dependent on the structure of the lattice. We define
the effective penetration depth of a vortex lattice by this equation using
the exact field distribution. This definition permits
a direct comparison between theory and experiment.  

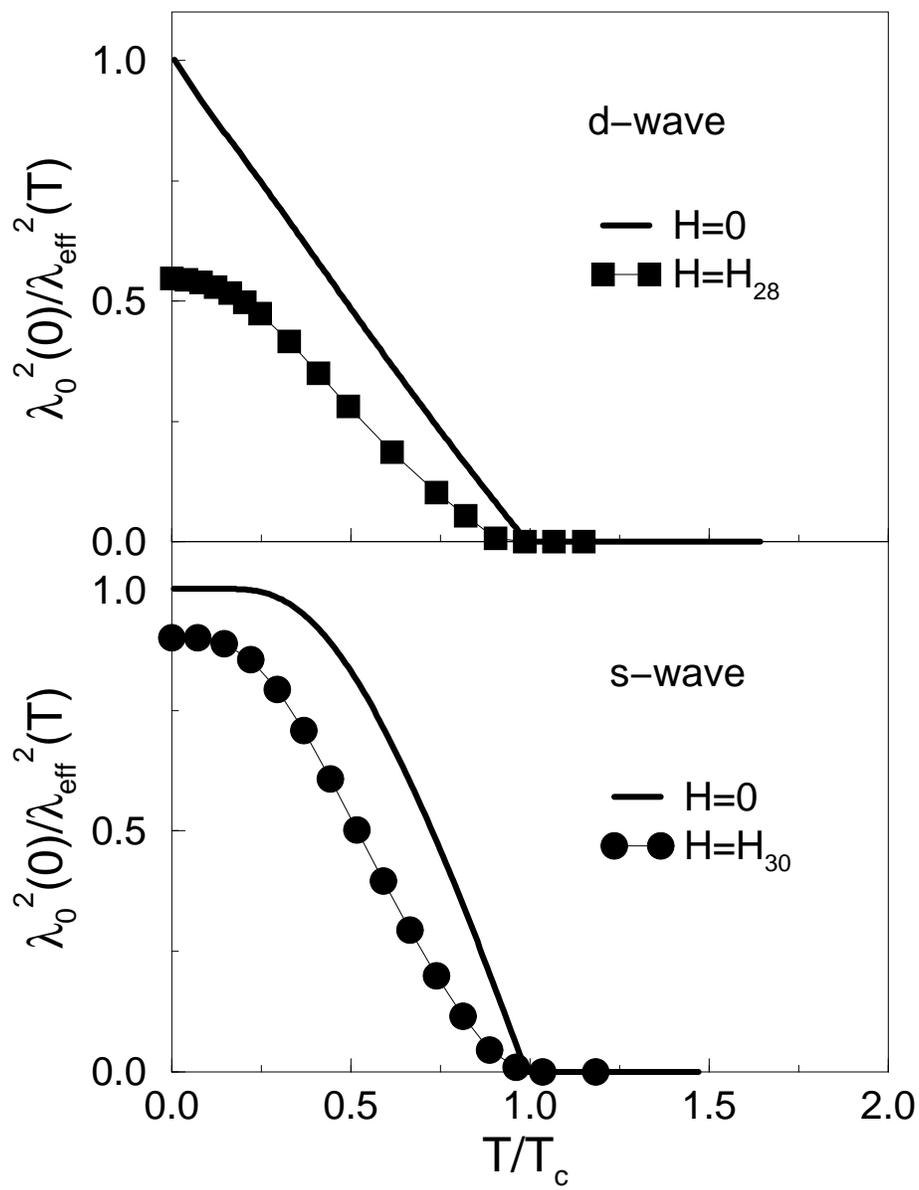
\begin{figure}
\caption{Temperature dependence of inverse square of the effective
penetration depth $\lambda_{eff}(T)$ in the mixed state ($H \simeq 0.04H_{c2}$)
for $d$-wave and $s$-wave superconductors, normalized by the zero-field,
zero-temperature penetration depth $\lambda_0(0)$, is plotted against
reduced temperature $T/T_c$.
$H_n$ represents the
field corresponding to the inter-vortex spacing of $na$.
For typical Cu-Cu distance in high $T_c$ materials,
$H_{30}$ and $H_{28}$, are 8.0 and 9.2 Tesla, respectively.
The zero-field penetration depth calculated with the same models are
shown for comparison.}
\label{fig:two}
\end{figure}

In Fig.~\ref{fig:two} we show the effective penetration depth 
$\lambda_{eff}$,  
as a function of temperature. We find that $\lambda_{eff}$   
increases with increasing magnetic field for both $d$-wave and 
$s$-wave superconductors, but much strongly in the $d$-wave case. Even in a 
relatively weak field, $H \simeq 0.04H_{c2}$, the 
low-temperature behavior in two
cases is very different. In the $s$-wave case, $\lambda_{eff}$   
essentially preserves an activated T-dependence. 
On the other hand, the linear temperature
behavior for the $d$-wave $\lambda$ in the 
Meissner state saturates at $T^{*}(H)$,
and crosses over to a $T^2$ dependence.  
The crossover temperature, $T^{*}$, can be crudely estimated by 
noting that the typical local pairing momentum in the vortex 
lattice state is $\sim \hbar /\ell$\cite{volovik}, 
where $2\pi \ell^2 H=\Phi_0$. This local superfluid 
velocity changes the quasiparticle density of states
\cite{sauls,stojkovic,volovik,yw_thesis} and hence 
$\lambda$ for energies below 
$k_B T^{*}(H) \sim v_F \hbar /\ell \sim \Delta_0 \sqrt{H/H_{c2}}$, 
where $\Delta_0$ 
is the maximaum quasipartice energy gap in the Meissner state.   
This estimate of the crossover temperature is consistent with the numerical
results shown in Fig.~\ref{fig:two} and similar results obtained at still 
stronger magnetic fields. 

Our calculation should be compared with 
recent $\mu$SR experimental data on YBa$_2$Cu$_3$O$_{6.95}$ reported by 
Sonier {\it et al.}\cite{kiefl}. 
These authors find that the penetration depth increases
with field. The increase of approximately $15\%$ they find at the 
strongest field studied, $H=2$ Tesla, should be compared with the 
approximately $30\%$ increase we find for $H/H_{c2}\simeq 0.04$. ($H_{c2}$
is belived to be $\sim 200$ Tesla for YBa$_2$Cu$_3$O$_{6.95}$). This is 
consistent with the expected approximate relationship between the 
zero-temperature penetration depth and the quasiparticle density of 
states at the Fermi energy which is 
$\propto \sqrt{H}$\cite{wang1,volovik,moler}.
The observed change in $\lambda_{eff}$ at this field is much larger than
would be expected for an $s$-wave superconductor, adding to the evidence
for $d$-wave pairing in cuprate superconductors. This experimental study 
has insufficient data at sufficient low temperature and 
sufficient strong fields 
to verify the crossover $T^2$ behavior predicted here.     


This work was supported by the National Science Foundation under
grant DMR-9416906.  The authors are grateful to 
A.J. Berlinsky, C. Kallin, R.F. Kiefl, M. Norman, J.E. Sonier, 
T.M. Riseman, and M. Franz for helpful interactions.

\begin{references}
\bibitem{latticestructure} We have obtained solutions for
both square and triangular vortex lattices.
The actual vortex lattice structure depends on detailed materials
\cite{mcmaster2,neutrons}. The physics discussed here is independent
of structure of vortex lattice.
\bibitem{remark1} Approximating the internal magnetic field appearing
in the BdG equations by a constant is valid in the extreme type-II
superconductors of interest, except at fields very close to the
lower critical field $H_{c1}$.
\bibitem{remark2} See Ref.~\cite{cincinatti}.
We expect that adding second neighbor hopping to our
Hubbard model would mitigate this tendency.
\bibitem{remark3} The dips along the diagonal directions result from nearly
square-shaped Fermi surface for a nearly half-filled band.
At smaller band-fillings
where the Fermi surface is close to a circle,
these dips are less pronounced.
\end{references}
\end{document}